\documentclass[%
 aip,jcp,
 prb,%
 amsmath,amssymb,
 reprint,%
]{revtex4-1}

\usepackage{dcolumn}
\usepackage{bm}

\usepackage{natbib}

\usepackage[usenames,dvipsnames,svgnames,table]{xcolor}
\usepackage{graphicx}
\usepackage[space]{grffile}
\usepackage{latexsym}
\usepackage{amsfonts,amsmath,amssymb}
\usepackage{url}
\usepackage[utf8]{inputenc}
\usepackage{microtype}
\usepackage{fancyref}
\usepackage{hyperref}
\hypersetup{colorlinks=false,pdfborder={0 0 0},}
\usepackage{textcomp}
\usepackage{longtable}
\usepackage{multirow,booktabs}

\newcommand{\remove}[1]{}

\begin{document}

\title{On the Consistency of Approximate Quantum Dynamics Simulation Methods
for Vibrational Spectra in the Condensed Phase}

\author{Mariana Rossi}
\affiliation{University of Oxford, South Parks Road, Oxford OX1 3QZ, UK}
\author{Hanchao Liu}
\affiliation{Emory University, 1515 Dickey Drive NE, Atlanta, GA 30322-1003, USA}
\author{Francesco Paesani}
\affiliation{University of California, San Diego, La Jolla, California 92093-0314, USA}
\author{Joel Bowman}
\affiliation{Emory University, 1515 Dickey Drive NE, Atlanta, GA 30322-1003, USA}
\author{Michele Ceriotti}
\affiliation{Laboratory of Computational Science and Modeling, IMX, {\'E}cole Polytechnique F{\'e}d{\'e}rale de Lausanne, 1015 Lausanne, Switzerland}

\date{\today}

\begin{abstract}
Including quantum mechanical effects on the dynamics of nuclei in the
condensed phase is challenging, because the complexity of exact methods
grows exponentially with the number of quantum degrees of freedom.
Efforts to circumvent these limitations can be traced down to two
approaches: methods that treat a small subset of the degrees of
freedom with rigorous quantum mechanics, considering the rest of the
system as a static or classical environment, and methods that treat the
whole system quantum mechanically, but using approximate dynamics. Here
we perform a systematic comparison between these two philosophies for
the description of quantum effects in vibrational spectroscopy, taking
the Embedded Local Monomer (LMon) model and a mixed quantum-classical
(MQC) model as representatives of the first family of methods, and
centroid molecular dynamics (CMD) and thermostatted ring polymer
molecular dynamics (TRPMD) as examples of the latter. We use as
benchmarks D$_2$O doped with HOD and pure H$_2$O at three distinct
thermodynamic state points (ice Ih at 150K, and the liquid at 300K and
600K), modeled with the simple q-TIP4P/F potential energy and dipole
moment surfaces. With few exceptions the different techniques yield 
IR absorption frequencies that are consistent with one another within 
a few tens of cm$^{-1}$. Comparison with classical molecular dynamics 
demonstrates the importance of nuclear quantum effects up to 
the highest temperature, and a detailed discussion
of the discrepancies between the various methods let us draw some
(circumstantial) conclusions about the impact of the very different
approximations that underlie them. Such cross validation between
radically different approaches could indicate a way forward to further
improve the state of the art in simulations of condensed-phase
quantum dynamics.
\end{abstract}

\maketitle

The simulation of nuclear quantum dynamics in the condensed phase remains a challenge even for state of the art methods. The main problem is that the dimensionality of these systems is too large for a fully ``first-principles'' quantum calculation of the nuclear dynamics, leaving the whole field without a reliable benchmark. Therefore, it is hard to assess how well different approximations perform even for exhaustively studied systems like bulk water. There is a dire need to find a reliable framework in which the quantum dynamics of these systems can be treated, so that the field can advance further.

One can identify at least two classes of popular methods to model quantum dynamics in the condensed phase. The first treats a subset of the system's degrees of freedom fully quantum mechanically, embedded in the electrostatic environment of the rest of the system. Examples of these methods are the local monomer approximation (LMon) \cite{Wang-LMon,Wang_2012}, which treats quantum mechanically a small subset of all intramolecular degrees of freedom in a static environment, and a mixed quantum-classical (MQC) model  based on the semi-classical theory of line shape, that treats a single degree of freedom (e.g. an OH stretch) ``on-the-fly'' in the presence of perturbing electric field generated by the surrounding classical solvent \cite{Torii_2006,Skinner_2009,Gruenbaum_2013}. The second class of methods treats the whole system on the same footings, but using (physically motivated) \textit{ad hoc} approximations to the quantum dynamics. These methods are often based on classical trajectories initiated from quantum distributions \cite{Miller_2005,Liu_2006,Liu_2009}, or are inspired by imaginary time path integral molecular dynamics (PIMD), as in the cases of centroid molecular dynamics (CMD)\cite{Cao_1994,Jang_1999} and (thermostatted) ring polymer molecular dynamics (TRPMD)\cite{Craig_2004,Habershon_2013,Rossi_2014}. Since these two classes of approaches are grounded on very different approximations, a comparison between their results is perhaps the closest one can get to assess their reliability in the absence of an absolute, exact benchmark \cite{Paesani_2010, AhlbornMoore_1999}. 

Here we perform such cross-validation, comparing the behavior of MQC, LMon, CMD, and TRPMD when modeling a physical observable that is particularly sensitive to the description of nuclear quantum dynamics -- the infra-red absorption spectra of hydrogen-containing compounds \cite{Habershon_2008,Witt_2009,Ivanov_2010,Paesani_2010,Rossi_2014}. We focused on prototypical hydrogen-bonded condensed phase systems -- HOD in D$_2$O and neat water -- describing them with the inexpensive q-TIP4P/F potential~\cite{Habershon_2009}, and using the corresponding linear dipole moment surface (DMS) to evaluate the IR absorption. The rationale for this simplistic choice is that we are not so much interested in comparing with experimental data, but only to nail down differences between methods for condensed-phase quantum dynamics. The simplicity of the potential and the linear DMS make it possible to guarantee thorough statistical sampling and to focus on the problem of modeling the dynamics on a complex, anharmonic potential energy surface, rather than on the fact that CMD and (T)RPMD are much harder to justify when computing correlation functions of non-linear operators \cite{JangVoth_2014, BraamsMano_2006, HeleAlthorpe_2013}. It also allows us to test extensively the dependence of TRPMD, CMD, and LMon to the precise details of the calculations. To validate the two philosophies over a broad range of conditions, and draw conclusions that are not purely anecdotal, we considered three very different thermodynamic regimes: ice Ih at 150K, liquid water at 300K, and the hot liquid at 600K at the experimental liquid/vapor coexistence density. To assess the importance of these approximate nuclear quantum dynamics, we also compare our results to classical MD simulations.

CMD and TRPMD simulations were performed using the i-PI code \cite{Ceriotti_2014}, and LAMMPS as the force back-end \cite{Plimpton_1995}. The details of the q-TIP4P/F potential are available elsewhere \cite{Habershon_2009}, as well as the technical details of the integration and thermostatting of PIMD \cite{Ceriotti_2010}.  We also performed classical MD reference calculations, using a time step of 0.5 fs and a weak global thermostat~\cite{Bussi_2007}. For both TRPMD and CMD we applied a Langevin thermostat to the internal modes of the ring polymer. In TRPMD we set the friction coefficients equal to the free ring-polymer frequencies: $\gamma_k=\omega_k$. We 
made this choice as a compromise between the damping of spurious peaks in the spectrum and the broadening of the physical peaks~\cite{Rossi_2014}, and used a time step of 0.25 fs.
In our CMD calculations we scaled the dynamical masses of the internal modes of the ring polymer to shift all the frequencies to the common value $\Omega=16000$cm$^{-1}$, and used a time step of 0.025fs, which makes CMD simulations ten times more demanding than TRPMD for the same level of statistics. CMD has better sampling efficiency than plain RPMD due to the thermostatting of the ring polymer normal modes~\cite{pere+09jcp}, but the efficiency of TRPMD should be comparable since it uses an analogous thermostat. 
In running the CMD calculations we noticed that the details of the thermostat have a significant impact on the centroid dynamics. As discussed in the Supplementary Material (SM)\cite{sm}, the commonly-adopted prescription of optimal coupling of either Langevin or Nos{\'{e}}-Hoover thermostats with the shifted frequency interferes with adiabatic decoupling at least up to $\Omega \leq 16000$ cm$^{-1}$ in the cases studied here. We found that in the limit of weak thermostat coupling ($\lambda=0.01$), proper adiabatic separation can be reached already at $\Omega<8000$ cm$^{-1}$, which could probably enable using longer time steps. This should be considered when performing CMD calculations and when comparing with existing results. 

All our LMon calculations were performed using the MULTIMODE code\cite{Carter_1997, Carter_1998, Bowman_2003}.
The LMon method assumes it is possible to take a snapshot configuration of a condensed-phase system, and solve exactly the Schr{\"o}dinger equation for a subset of its degrees of freedom, evaluating the changes of the potential as a function of deformations of a molecule along a subset of its ``lormal'' modes. A detailed explanation of this approach can be found for instance in Ref.\cite{Wang-LMon}. 
The LMon approximation makes it possible to progressively include more physics by increasing the dimensionality of the quantum subspace.  Using just the three intra-molecular ``lormal'' modes as the quantum-mechanical subspace (LMon-3) is generally sufficient to describe the stretching and bending regions of the spectrum of water\cite{Liu-Hexamer}. A recent study of vibrational energy relaxation (VER) of dilute HOD in ice Ih considered also three additional intermolecular modes (LMon-6)\cite{Liu_2014}, obtaining a description of the intermolecular vibrational spectrum and an  improved accuracy in the intra-molecular region, but at a much higher computational cost.
Here we are mostly interested in the strongly-quantized intra-molecular vibrations, and so we attempted a simplified approach that uses a total of four modes (LMon-4). Hundreds of monomer+environment configurations were obtained from individual replicas of a series of PIMD simulations. For each snapshot, the three intramolecular modes are complemented with the three highest-frequency inter-molecular modes, that are however considered one at a time: for each monomer, 3 sets of LMon-4 calculations are performed separately.  A smooth spectrum is obtained by combining Gaussian line shapes centered around the energies of transitions from the ground state to each of the excited states, weighed by the transition dipole moment. Further numerical details about all the calculations can be found in the SM.

MQC calculations were performed with an in-house code. Following Ref~\cite{Paesani_2010}, several independent CMD trajectories were performed starting from equilibrated PIMD configurations, from which the time-dependence of the vibrational frequency and transition dipole moment for the OH stretch were computed.  The semiclassical vibrational line shape was then evaluated as discussed in Ref.~\cite{mukamel1999principles}, including non-Condon corrections as defined, e.g., in Ref. \cite{Schmidt_2005}. A more detailed description of the simulations, as well as a discussion of the importance of non-Condon corrections, is reported in the SM.
From the point of view of computational cost, both MQC and LMon-4 are relatively inexpensive, the most demanding aspect being the preliminary (PI)MD simulation. Traditionally the environment is sampled classically -- which particularly at low temperature can be orders of magnitude less demanding -- but as we will see results are influenced by the classical or quantum treatment of the environment. Then, the evaluation of the stretching frequency in MQC is trivial, for such a simple potential.  Each LMon-3 calculation takes a few seconds on a single core, and about one minute for LMon-4. The cost would increase steeply as a function of the size of the quantum subspace, but up to LMon-6 the cost will most likely be dominated by the preliminary sampling of bulk configurations. 

Let us start by considering a single HOD molecule solvated in D$_2$O. This system (and its counterpart, HOD in H$_2$O) is commonly used as a probe for studying the local structure of liquid water theoretically and experimentally \cite{bakker2009vibrational,Paesani_2010}. The OH stretch is dynamically uncoupled with the environment, making this system well-suited for both MQC and LMon calculations~\cite{Liu_2014}. 

\begin{figure}[tbhp]
\centering
\includegraphics[width=0.82\columnwidth]{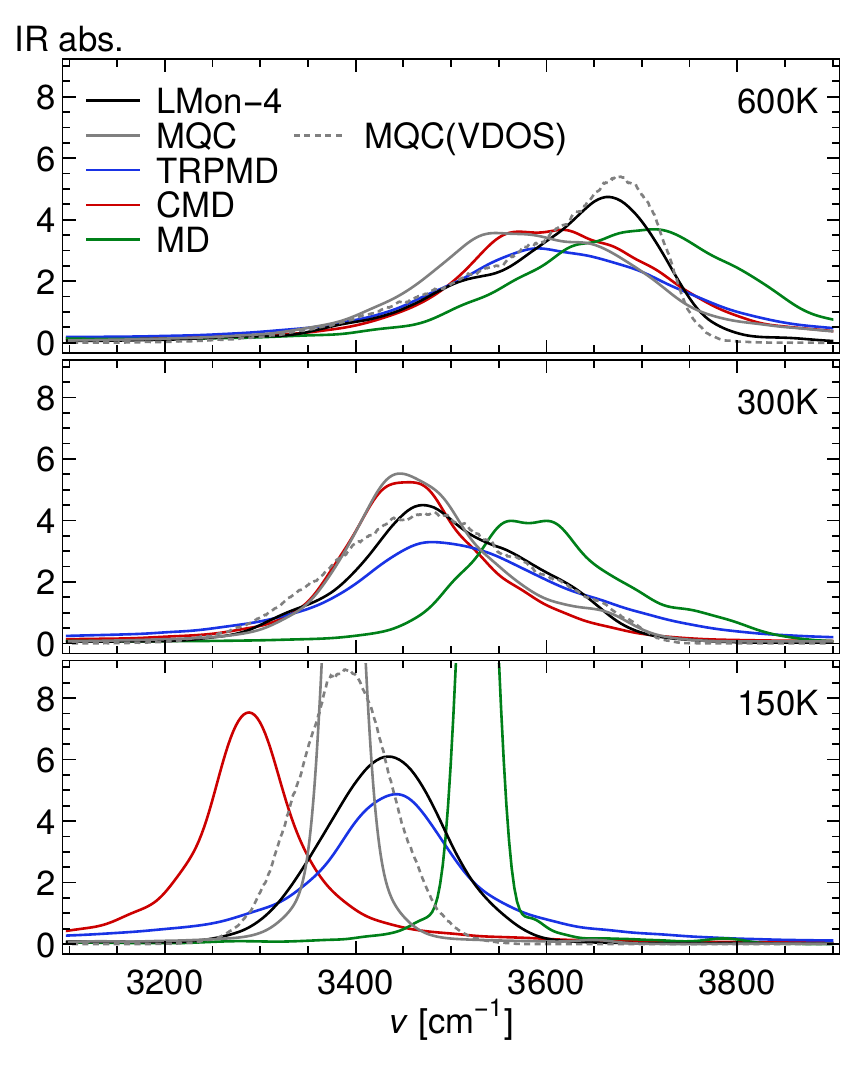}\vspace{-4mm}
\caption{\label{fig:hod-d2o} Comparison between the OH stretch IR absorption spectrum 
for a single HOD molecule in bulk D$_2$O, modeled using the q-TIP4P/F potential. 
Absorption spectra were computed from the dipole derivative autocorrelation, 
using classical molecular dynamics (green), CMD (red), and TRPMD (blue), and compared 
with the results of LMon-4 calculations (black) and MQC line shape (gray) and 
VDOS (gray dotted). The panels correspond, from top to bottom, to liquid water at 
600 K, liquid water at 300 K and ice Ih at 150 K. The integrated intensity 
of the OH stretch peak has been normalized to the same area.}
\end{figure}

The results for classical MD, CMD, TRPMD, MQC and LMon are shown in Figure \ref{fig:hod-d2o}. Starting from the highest temperature, 600 K, with a hot and compressed liquid, we observe that classical MD is blue-shifted by around 50 cm$^{-1}$ with respect to all quantum methods. Even at this high temperature nuclear quantum effects affect the calculation of dynamical properties. The blue shift of the classical simulation gets more pronounced by lowering the temperature, as expected, amounting to approximately 100 cm$^{-1}$ at 150K. 
At 600 K all quantum methods show a large blue-shift, increased band width and asymmetric or structured line shapes, compared to those at 150 and 300 K. The LMon-4 band intensity falls-off faster than other line shapes at the high-frequency edge of the band, and shows a sharper maximum which is 30-50 cm$^{-1}$ to the "blue" of the CMD and TRPMD maxima, which are not sharp. Considering that the influence of low-frequency intermolecular modes, e.g., hot bands, additional dipole variation, vibrational relaxation, etc.,  becomes more important with increasing temperature and that LMon-4 is only partially describing this coupling, the limitations in the LMon- 4 theory are also expected to be become more apparent. The importance of inter-molecular couplings and of the dynamics of the environment at the higher temperature is also suggested by the large difference we observe between the MQC line shape, the Condon approximation to the MQC line shape, and MQC vibrational density of states (VDOS, see the SM).

At 300 K the agreement between different quantum techniques is perhaps even better. MQC and CMD overlap almost perfectly. The LMon peak is some 20-40 cm$^{-1}$ higher in frequency, and the difference between the LMon and MQC line shapes is much less dramatic than at 600K. At 300 K, inhomogeneous broadening effects (i.e., effects \textit{not} related to vibrational relaxation or motional narrowing) are dominant. The TRPMD peak is further blue-shifted (by 20-40 cm$^{-1}$, depending on whether one considers the maximum or the mean position of the peak) and artificially broadened by the strong thermostatting of non-centroid modes.

In ice Ih at 150K TRPMD and LMon still are in remarkable - although perhaps fortuitous - agreement.
CMD shows a pronounced red shift of 180 cm$^{-1}$, which should probably be attributed to 
the curvature problem \cite{Witt_2009,Rossi_2014}. 
The MQC peak is red-shifted by about 60 cm$^{-1}$ relative to LMon and TRPMD. 
One possible explanation for this shift is the description of the environment based on centroid 
configurations, that are very close to classical. Contrary to higher temperatures, the LMon spectrum 
also depends on whether the environment configurations are obtained from the beads 
of the PIMD simulation, or from the centroid. 
In the latter case, as when using configurations from classical MD,
the OH stretch peak is red-shifted by 25 cm$^{-1}$ (see the SM). 
It is arguable which choice is more physically justified. On one hand, bead positions provide 
a statistically accurate snapshot of the quantum environment. On the other, 
centroids can be seen as a mean-field average of the quantum fluctuations of the neighboring 
molecules, closer perhaps to the spirit of a quantum-classical model. In the absence of a rigorous 
justification we can see this discrepancy as a sign of the break-down of the classical model 
for the environment, and consider the difference between the two spectra as an estimate of 
the reliability of LMon (and MQC) in this low temperature regime. 

\begin{figure}[tbhp]
\centering\includegraphics[width=0.82\columnwidth]{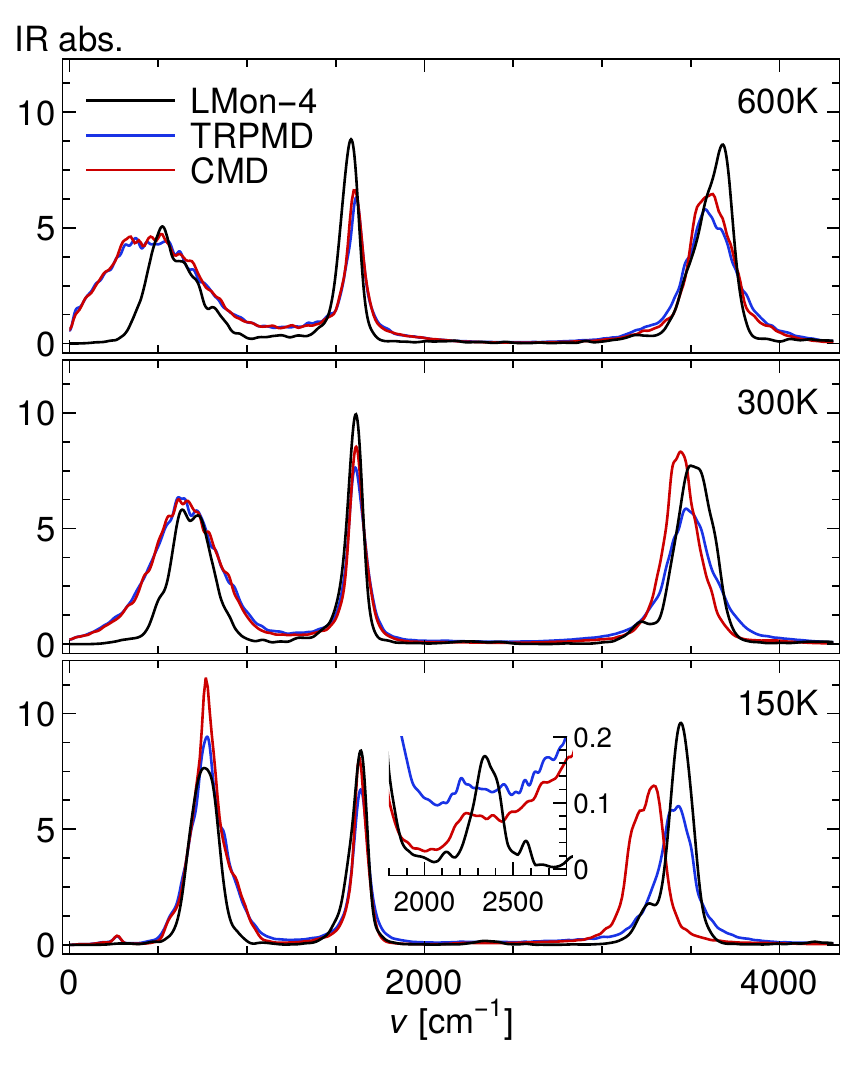}\vspace{-5mm}
\caption{\label{fig:water} Infra-red absorption spectrum of H$_2$O at three different thermodynamic conditions.
From top to bottom: (compressed) liquid water at 600 K, liquid water at 300 K and ice Ih at 150 K. 
The curves correspond to TRPMD (blue), CMD (red) and LMon-4 calculations (black). 
Note in the middle panel the reduced red shift of the OH peak in CMD compared with
 the results in Ref.~\cite{Rossi_2014}. The coupling of efficient thermostats with the internal 
modes of the ring polymer, that was used in the previous work, caused a spurious red shift of 
about 30 cm$^{-1}$ (see also the SM).}
\end{figure}

Having compared all methods for the HOD:D$_2$O benchmark, we consider in Fig. \ref{fig:water} 
the spectra of water in the three thermodynamic regimes discussed above. We did not perform 
MQC simulations, since the isolated chromophore assumption is less justified in H$_2$O.
Let us start by discussing the OH band, that shows similar trends to those observed for 
HOD in heavy water. At 600K, CMD and TRPMD are in near-perfect agreement, and the peak 
maxima are slightly red-shifted relative to LMon-4 by about 50 cm$^{-1}$. This is 
consistent with our observations in HOD case, and the discrepancy can be attributed to the 
lack of homogeneous broadening and dynamical couplings in LMon. At 300K TRPMD and LMon-4 
are close to each other, while CMD shows a small red shift of less than 50 cm$^{-1}$, 
consistently with what observed in HOD:D$_2$O. In the case of low-temperature ice the CMD 
peak is red shifted by 150 cm$^{-1}$  compared to both LMon-4 and TRPMD, that agree 
well with each other -- even though the TRPMD peak is considerably broader. 
The intensity of the stretching band is much under-estimated relative to experiments.
This is due to the linear DMS of q-TIP4P/F, as evidenced by comparison with 
more sophisticated models~\cite{Wang_WHBB}, but is irrelevant for our 
comparison of approximate methods for quantum dynamics.

Moving on to the bend, we observe good agreement in the peak position and width 
among all the quantum methods. All predict a slight red shift with increasing 
temperature (in qualitative agreement with experiment).  TRPMD and CMD give peaks 
positions at  1640 cm$^{-1}$ at 150 K and about 1610 cm$^{-1}$ in at 600 K.
 LMon-4 gives 1640 cm$^{-1}$ at 150 K and 1590 at 600 K.  The peak at 600K is 
further lowered to 1570cm$^{-1}$ when using LMon-3 (see the SM), indicating the 
growing importance of inter-molecular coupling at high temperature, and suggesting 
that this slight discrepancy between LMon-4 and PIMD-based methods could be resolved 
by increasing further the dimensionality of the quantum subspace.
Finally, in the low-frequency region CMD and TRPMD are almost identical. In this 
region one cannot expect LMon-4 to yield quantitative accuracy.  The quantum sub-space 
does not contain the collective modes of the hydrogen-bond network, nor the 
translation modes of individual monomers.

Within the LMon scheme it is straightforward to treat effects beyond linear 
absorption, and to give a clear physical attribution of specific features of the spectrum. 
For instance, all of the spectra display a distinct bump or shoulder around 3200 cm$^{-1}$, 
which corresponds to the first overtone of the bend. An interesting feature captured by 
LMon-4 (but not by LMon-3) is the small peak around 2350 cm$^{-1}$, which is evident
at 150K and becomes less clear-cut with raising temperature.
This feature, is due to the combination band of bending and librational modes, 
and demonstrates how increasing the dimensionality of the quantum subspace 
rogressively includes additional physical effects in LMon calculations. 
Both CMD and TRPMD display non-zero absorption in 
this region, but further analyses would be needed to attribute that spectral density 
to a precise physical origin. 

The conclusions of our comparison of approximate quantum dynamics simulation methods 
are overall optimistic: all the methods we considered are generally consistent with each other,  
while the difference with the position of the OH stretching peak observed in classical 
molecular dynamics confirms the importance of including nuclear quantum effects 
to reproduce quantitatively spectroscopic measurements of hydrogen-containing systems.
 TRPMD and LMon agree within a few tens of cm$^{-1}$ over a 
range of thermodynamic conditions going from ice Ih at 150K to water at 300K and finally 
to the hot, compressed liquid at 600K. So do CMD down to room temperature, and MQC in the 
cases where it is applicable. Furthermore, this analysis reinforces the notion that 
performing PIMD-based and quantum-subspace simulations in tandem does not only 
provide a degree of cross-validation, but also makes it possible to profit simultaneously 
from the complete (albeit approximate) description of the absorption spectrum given by 
the former family of methods, and from the interpretation of distinct spectral 
features that is enabled by an explicit quantum treatment\cite{Paesani_2009}.

The comparison between the different methods also highlights their current limitations, 
and suggests that results of quantum dynamics should always be interpreted with caution. 
Extensions of MQC beyond the case of an isolated chromophore require one to model 
the interaction between the different chromophores, either explicitly or by devising 
an effective mapping of the molecular Hamiltonian on a collective coordinate, such as 
the local electrostatic potential\cite{ChoiCho_2013}. Even though LMon can in principle be
improved systematically, its computational cost increases exponentially with the 
size of the quantum subspace, making it challenging to fully assess the 
importance of inter-molecular couplings. Treating the environment semiclassically, 
in the same spirit as the MQC model, might be a better way to account for inter-molecular 
dynamical effects. One has also to consider that the present work represents the first 
application of LMon to a finite-temperature scenario. More work is needed to include 
systematically a description of monomer-monomer couplings, hot bands and the dynamical 
nature of the environment. Particularly at the highest temperature, these effects may 
affect the line shape of the absorption peaks, as highlighted by the comparison with 
MQC. We also observe that at 150K the results of a LMon 
calculation depends sizably on whether one takes snapshots of the environment that are 
consistent with quantum statistics (e.g., from the beads in an PIMD simulation) or 
that are essentially classical (e.g., taken from classical MD, or using the ring polymer centroids). 

Methods inspired by PIMD also have their issues. In its straightforward 
partially-adiabatic implementation, CMD requires a small integration time step, 
and one should pay attention to the fine details of the thermostatting
to avoid interfering with the adiabatic decoupling of the centroid. 
At low temperatures, a pronounced red-shift relative to all other methods is observed. 
The precise temperature where this artifact becomes sizable should depend on the details 
of the system being studied. TRPMD, on the other hand, can be straightforwardly used 
with the same time step of a PIMD calculation, and does not suffer from systematic 
low-temperature red shifts. However, the OH stretch mode can be up to 50cm$^{-1}$ blue-shifted 
relative to some of the other techniques. Furthermore, the precise peak position varies depending 
on the choice of the thermostat coupling, which is somewhat arbitrary. 
The strong thermostatting that is used to fix the resonance problems of RPMD also leads to a 
significant broadening of the peaks, so the TRPMD line shape should not be taken too seriously, 
unless it is dominated by inhomogeneous broadening or is weakly quantized. 

Our results give some confidence on the reliability of approximate quantum dynamical 
methods for simulating the dynamics and the spectroscopy of condensed-phase systems or 
large molecules. The difference between quantum and classical simulations is comparable 
to the errors due to imperfect potentials, but the discrepancies between different quantum 
methods are less important. The best agreement between the various methods is seen 
for liquid water at room temperature -- perhaps unsurprisingly considering that it corresponds 
to the regime for which most of approximate quantum techniques were designed. Cross-validation 
between different approaches, particularly at thermodynamic conditions outside of their
 ``comfort zone'', offers the best promise for demonstrating the accuracy of existing 
methods, for uncovering their errors and shortcomings, and for delivering more reliable, 
simple and affordable simulation strategies.

\vspace{-5mm}
\section{Acknowledgements}
\vspace{-3mm}We gratefully acknowledge stimulating discussion with David Manolopoulos, Gregory Voth, and Lu Wang. MR thanks the Deutsche Forschungsgemeinschaft (DFG), RO 4637/1, for funding. HL and JMB thank the US National Science Foundation, CHE-1145227, for funding. FP thanks the US National Science Foundation, CHE-1111364, for funding, and the Extreme Science and Engineering Discovery Environment (OCI-1053575, TG-CHE110009) for computational time.

\setcounter{figure}{0}
\renewcommand\thefigure{S\arabic{figure}}    
\appendix
\section{Details of TRPMD and CMD simulations}

It is well known from Fermi's golden rule that the IR spectrum can be obtained by \cite{mcquarrie} 

\begin{eqnarray}
n(\omega)\alpha(\omega) = \frac{\pi \omega}{3 \hbar c V \epsilon_0}(1-e^{-\beta \hbar \omega})I_{\mu\mu}(\omega) \\
I_{\mu\mu}(\omega) = \frac{1}{\pi}\int_0^\infty e^{i\omega t}\langle \bm{\mu}(0) \cdot \bm{\mu}(t) \rangle dt
\end{eqnarray}

\noindent where $\mu$ is the dipole moment, $n(\omega)$ the refractive index, $\alpha(\omega)$ the Beer-Lambert absorption coefficient, and $V$ the volume of the simulation box.  The angular brackets indicates a canonical average. TRPMD and CMD can be better described as approximations to the Kubo-transform time correlation, which is given by \cite{Habershon_2008}

\begin{eqnarray}
\tilde{c}_{\mu\mu}(t)=\frac{1}{\beta} \int_0^\beta d\lambda \langle \bm{\mu}(-i\lambda\hbar) \cdot \bm{\mu}(t) \rangle \label{eq:kubo}
\end{eqnarray} 

\noindent that has a simple Fourier transform relation to the canonical autocorrelation function given by

\begin{eqnarray}
\tilde{I}_{\mu\mu}(\omega)=\frac{(1-e^{-\beta \hbar \omega})}{\beta \hbar \omega}I_{\mu\mu}(\omega)
\end{eqnarray} 

\noindent where $\tilde{I}_{\mu\mu}(\omega)$ is the Fourier transform of Eq. \ref{eq:kubo}. The final expression is, thus

\begin{eqnarray}
n(\omega)\alpha(\omega) \propto \omega^2 \tilde{I}_{\mu\mu}(\omega) \label{eq:IR-spec}.
\end{eqnarray}

Instead of using directly the equation above to calculate our spectra with the Fourier transform of the dipole autocorrelation, we have
instead calculated the Fourier transform of the autocorrelation of the dipole \textit{time derivative}. These two quantities are related by

\begin{equation}
\tilde{I}_{\dot{\mu}\dot{\mu}}(\omega) = \omega^2 \tilde{I}_{\mu\mu}.
\end{equation}

The time derivative was taken numerically from two consecutive time steps and the final expression we used for
the IR spectra shown in the paper is thus

\begin{eqnarray}
n(\omega)\alpha(\omega) \propto \tilde{I}_{\dot{\mu}\dot{\mu}}(\omega).
\end{eqnarray}

Since the dipole is only a function of positions $\mathbf{q}$, in the path-integral based approximations, the estimator for the dipoles is given by $\bm{\mu} = 1/P \sum_{i=1}^P \bm{\mu}_P(\mathbf{q})$, where $i$ runs through the $P$ replicas (beads) of the ring polymers. In the q-TIP4P/f model used here the dipole is linear with respect to the positions, such that we could take simply the dipole moment of the centroid configurations in the expression above. 

Given the difficulty of directly comparing the spectra coming from all different approaches treated in the manuscript, we have normalized the integrated intensity of all spectra to be the same.

We performed initial equilibration runs of 100 ps in order to obtain starting structures for the TRPMD and CMD simulations for all system at each temperature studied. For HOD:D$_2$O ice at 150K we performed 8 independent initial runs with H in different positions in order to account for proton disorder. We disregarded the first 5 ps of each thermalization run and took starting geometries (5 to 10 ps apart) from the rest of these simulations. Starting geometries for CMD and TRPMD were taken from the same thermalization runs in each case, and several independent runs were performed.  In total, for HOD:D$_2$O we ran 20 ns of simulations for TRPMD and 6 ns for CMD for each temperature -- a very long time to ensure the convergence of the spectra, which is especially difficult in this case since we could not exploit horizontal statistics. We calculated the HOD IR absorption spectrum from the dipole of the HOD molecule only. For H$_2$O we performed 100ps of TRPMD and 100ps of CMD for each temperature, since here we could exploit horizontal statistics. We obtained the IR spectrum from the total dipole moment of the cell.

The time step used for these simulations was 0.5~fs for MS, 0.25fs for TRPMD and 0.025fs for CMD. In all cases, to improve ergodicity, we attached to the atoms or the centroid 
a weak global thermostat that has been proven not to affect dynamical properties~\cite{Bussi_2007}.
The number of beads used for ice Ih at 150 K was 64, and we took a box containing 96 water molecules at the experimental structure and density. For the liquid at 300 K we used 32 beads and a box containing 128 water molecules, also at experimental density. For the high temperature simulation at 600K we used 16 beads, a box of 128 water molecules, and the experimental density at the liquid-vapor coexistence at this temperature. 

For the Fourier transform of the dipole autocorrelation series we used a lag of 1 ps and padded  the signal with zeros in order to
obtain a grid of around 3 cm$^{-1}$ in frequency space. We also windowed the autocorrelation function with a Hanning window in order to filter spurious ringings of the signal.

\section{Details of LMon calculations}

As discussed in the text, we opted here for a compromise on the size of the quantum subspace, using the LMon-4 as an effective approximation to LMon-6. For each monomer, 3 sets of LMon-4 calculations are performed separately. Each LMon-4 calculation yields a series of energy states, and the corresponding matrix of transition dipole moments. Since we consider each snapshot three times, using a different inter-molecular mode but the same intramolecular modes, the approach lead to an over-estimation of the intensity of intramolecular absorptions. To avoid this overcounting we weighted all the transitions from the ground state to intramolecular excited states by a factor of 1/3. Transitions to states with non-zero quanta in the intermolecular mode correspond to different quantum mechanical states in each of the three LMon-4 calculations, so their intensity has been weighed in full. 

For dilute HOD in D$_2$O water, only the HOD monomer was consider for the LMon-4 calculations. In the case of ice we took configurations from eight independent realizations of the lattice to sample different electrostatic environments for the HOD molecule. For pure H$_2$O water, 5 monomers are sampled from each snapshot. The snapshots are between 0.5 and 1 ps apart from each other in order to ensure a better statistical sampling. The total numbers of monomer configurations for HOD are 2185, 1132 and 1165 for 150K, 300 K and 600 K respectively. For H$_2$O, the numbers are 2400, 1951 and 1486 for 150K, 300 K and 600 K respectively. The final spectra were smoothed with a Gaussian function with FWHM=50 cm$^{-1}$. This amplitude is much smaller than the intrinsic bandwidth of the features of the spectrum of water, and we only used it to eliminate the noise due to the use of a finite number of monomer configurations. 

As discussed in the main text -- for the relatively small dimensionality of quantum subspace we used -- the main bottleneck is the preliminary sampling of bulk configurations. This is particularly true when performing such sampling using path interal molecular dynamics, that depending on the implementation and the temperature might be orders of magnitude more demanding than classical molecular dynamics.

\begin{figure*}[tbhp]
\begin{center}
\includegraphics[width=0.7\textwidth]{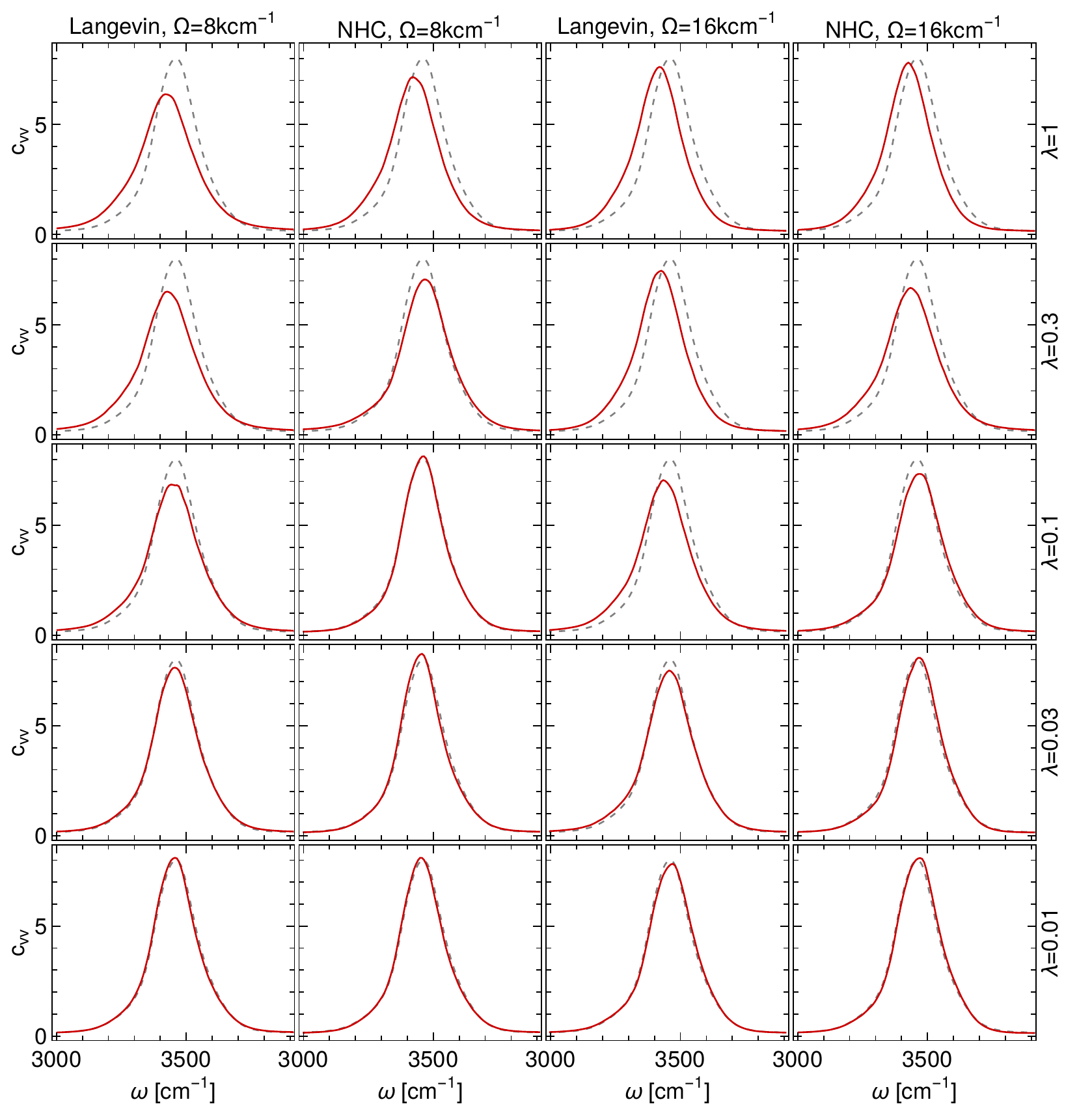}
\end{center}
\caption{\label{fig:vv-centroid}
OH-stretching peak for the centroid velocity correlation spectrum in a PA-CMD simulation of liquid water at room temperature. The different panels correspond to different NM dynamical frequency $\Omega$ and thermostat strength $\lambda$. The dashed line correspond to an average of the peaks obtained with $\lambda=0.01$, and is plotted as to guide to the eye.}
\end{figure*}

\begin{figure*}[tbhp]
\begin{center}
\includegraphics[width=0.7\textwidth]{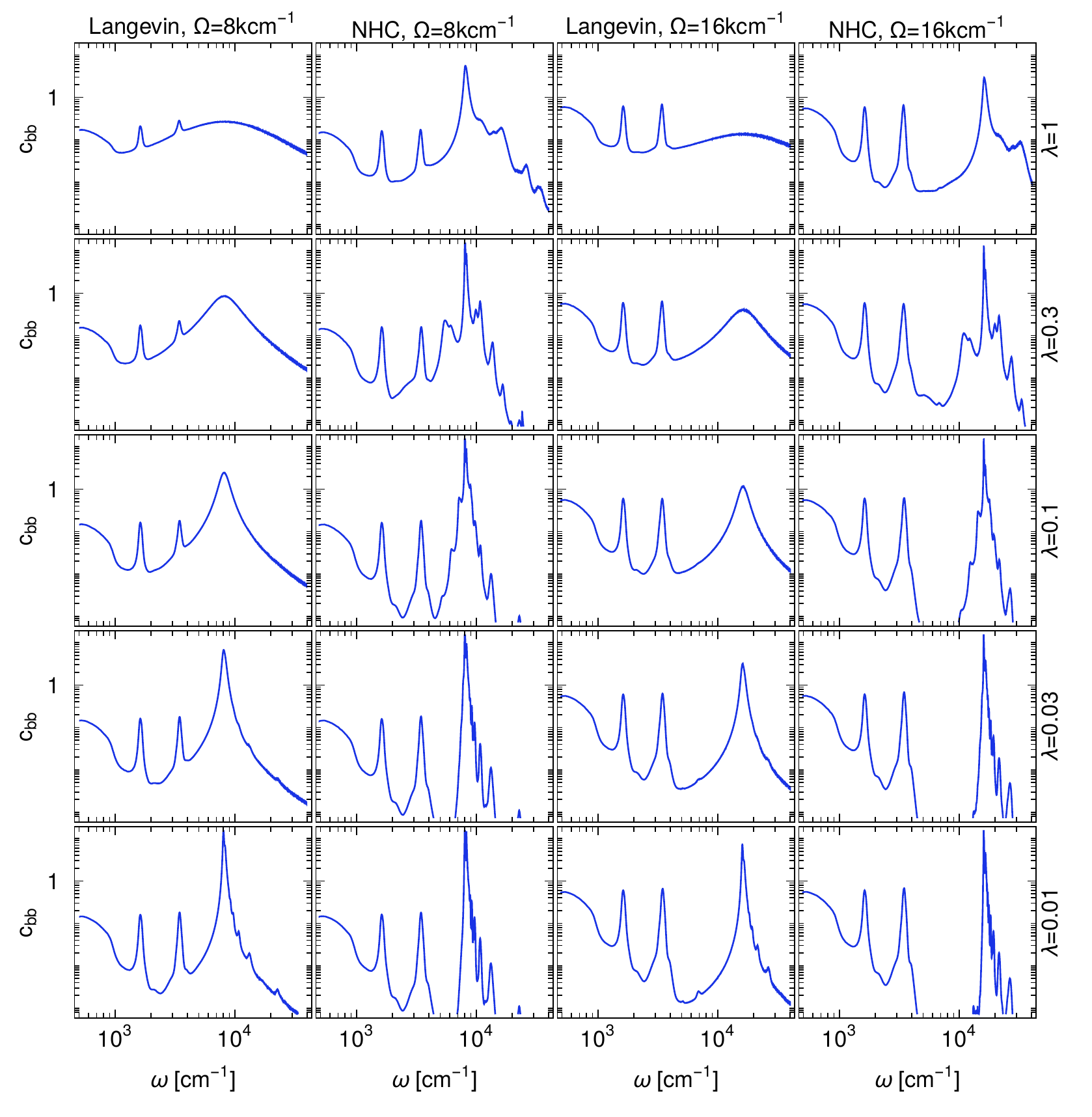}
\end{center}
\caption{\label{fig:vv-beads}
Velocity-velocity correlation function for one of the beads in a PA-CMD simulation of liquid water at room temperature. The different panels correspond to different NM dynamical frequency $\Omega$ and thermostat strength $\lambda$. The plots are reported on a log-log scale, to visualize clearly the different time scales involved. }
\end{figure*}

\section{Assessing the effect of normal-modes thermostatting in CMD}

Centroid molecular dynamics is defined as the (microcanonical) dynamics of the ring polymer centroid on the potential of mean force obtained by averaging the ring polymer Hamiltonian over the possible fluctuations of the beads with fixed centroid position~\cite{Cao_1994}. 

In practical implementations, one does not average explicitly over the internal modes of the ring polymer, but performs a simultaneous dynamics of the centroid and the non-zero frequency normal modes (NM) of the free ring polymer, defining the dynamical masses in such a way that (for a free particle) all the normal modes vibrate at a common frequency $\Omega$. By setting this frequency to a value much larger than any physical vibration in the system, one obtains a (partial) adiabatic decoupling between the centroid and the other NM (PA-CMD), i.e. the ring polymer potential energy surface is sampled dynamically as the centroid moves on a much slower time scale~\cite{Hone_2004}. 

As a consequence of this rescaling of the dynamical masses, one has to use smaller time steps to integrate the equations of motion -- that get smaller the larger the maximum dynamical frequency is. It is worth noting that the time scale for the integration is \emph{not} determined by the value of $\Omega$ alone, but by a combination of $\Omega$, inverse temperature $\beta$, and the maximum physical frequency present in the system, $\omega_\text{max}$. In fact, before the mass scaling, the slowest internal mode of the free ring polymer has a frequency of approximately $2\pi/\beta\hbar$. In the presence of an external harmonic potential with frequency $\omega_\text{max}\gg 2\pi/\beta\hbar$, its intrinsic vibrational frequency shifts to approximately $\omega_\text{max}$. Since the mass-scaling acts on all the physical degrees of freedom, and scales all the ring polymer vibrations by the same factor, irrespective of the underlying physical potential, a PA-CMD simulation of the system described above will contain a vibrational mode whose dynamical frequency is approximately $\Omega \omega_\text{max}\beta\hbar/2\pi$. For moderately quantum systems this is much larger than $\Omega$ itself.

The choice of the NM frequency $\Omega$ is generally understood to be crucial to obtain convergence of the dynamical properties computed by PA-CMD to the fully adiabatic values, while maintaining a compromise between the converged result and the use of a reasonably large time step.
Any work based on PA-CMD quotes clearly the value of $\Omega$, or an equivalent adiabaticity parameter that conveys the same information~\cite{Hone_2004}. However, if PA-CMD is meant to perform an on-the-fly average of the ring-polymer potential, the NM have to be properly thermostatted. The common wisdom is then to use a thermostat that is optimally coupled to the dynamical frequency $\Omega$, in order to sample the potential of mean force as efficiently as possible. For instance, when using a Nos{\'e}-Hoover chains thermostat~\cite{Hoover_1985,Nose__1984,Martyna_1992}, one should set the thermostat mass to $Q=P/\beta\Omega^2$~\cite{Kinugawa_1997}. When using a Langevin thermostat one should set the friction to the critical damping $\gamma=2\Omega$\cite{Ceriotti_2010}. 

Having observed difficulties in converging PA-CMD results with respect to $\Omega$ while using an optimally-coupled Langevin thermostat, we have performed an extensive investigation of how the centroid velocity correlation spectrum computed by PA-CMD depends separately on $\Omega$ and on NM thermostatting, for a simple flexible water model~\cite{Habershon_2009}, at a temperature of 300K, using 32 beads and a time step of 0.025fs. We describe the intensity of the coupling by a parameter $\lambda$, that is defined as $\lambda=\gamma/2\Omega$ for Langevin thermostatting, and as $\lambda^2=P/\beta Q\Omega^2$ for Nos{\'e}-Hoover thermostatting (chains of four thermostats, one chain per degree of freedom). In both cases, the parameter corresponds to optimal coupling to a scaled frequency $\lambda\Omega$, so that smaller $\lambda$ corresponds to weaker coupling. 

Figure~\ref{fig:vv-centroid} shows the results of this analysis. The amplitude and position of the OH stretching peak depends significantly on the NM thermostatting, both for Langevin and Nos{\'e}-Hoover thermostats. As $\lambda$ decreases, the peaks shift progressively to higher frequency, until converged results are reached in a weak coupling regime with $\lambda\approx 0.01$. This turns out to be the proper adiabatically-decoupled limit, and the peak position and shape does not change significantly if $\lambda$ is further decreased, or $\Omega$ further increased. This effect is very insidious, since in the limit of a strong thermostat the position changes very little when $\Omega$ is increased, which may give the impression that adiabatic decoupling has been reached. In fact, recent results by some of the Authors were obtained with $\Omega=16000$cm$^{-1}$ and $\lambda=0.2$, and were therefore red-shifted by about 30cm$^{-1}$ relative to the correct fully adiabatic limit -- thereby exaggerating the curvature problem~\cite{Rossi_2014}.

The origin of this strong thermostat dependence can be understood by looking at the overall dynamical behavior of the thermostatted ring polymer. Figure~\ref{fig:vv-beads} shows the bead velocity correlation spectrum, that displays signatures of the motion of all the ring polymer degrees of freedom simultaneously. One can notice clearly a large peak corresponding to the target NM frequency $\Omega$ and also, for weak couplings, the presence of additional peaks at much higher frequency that correspond to the combination of the physical potential with the spring terms in the path integral Hamiltonian. 

However, the NM vibrations are altered by the presence of a strong thermostat, that broadens the vibrational peak and causes spurious interactions between the internal modes and the centroid. This is reminiscent of the effect of a thermostat in Car-Parrinello molecular dynamics, where the use of a white-noise Langevin thermostat disrupts the adiabatic decoupling between the ionic and electronic degrees of freedom~\cite{Ceriotti_2009}. This effect is very evident in the case of Langevin thermostatting, while the impact of NHC thermostats on the dynamics is more subtle. 
The fact that NHC thermostats alter the peak shape nearly as much as Langevin dynamics can be probably be understood by the fact that NH equations are non-linear, and therefore not invariant to unitary transformations of the coordinates~\cite{Ceriotti_2010}, and so they lead to stronger coupling between centroid and non-centroid modes than one would imagine based on the effect on the velocity correlation spectrum.

When using optimally-coupled thermostats, the strength of the disturbance of the dynamics increases together with $\Omega$, and so it would take a huge adiabatic decoupling to eliminate the influence of the thermostat. On the other hand, using a weaker coupling reduces dramatically the effect of the the thermostat, so that satisfactory adiabatic decoupling can be reached with smaller values of $\Omega$. In fact, there have been cases in which \emph{no} thermostat was attached to the internal modes of the ring polymer~\cite{Habershon_2008,scot-manoprivatecommunication}, which made PA-CMD converge with very weak adiabatic decoupling, without apparent adverse effects. 

\begin{figure}[tbhp]
\begin{center}
\includegraphics[width=0.7\columnwidth]{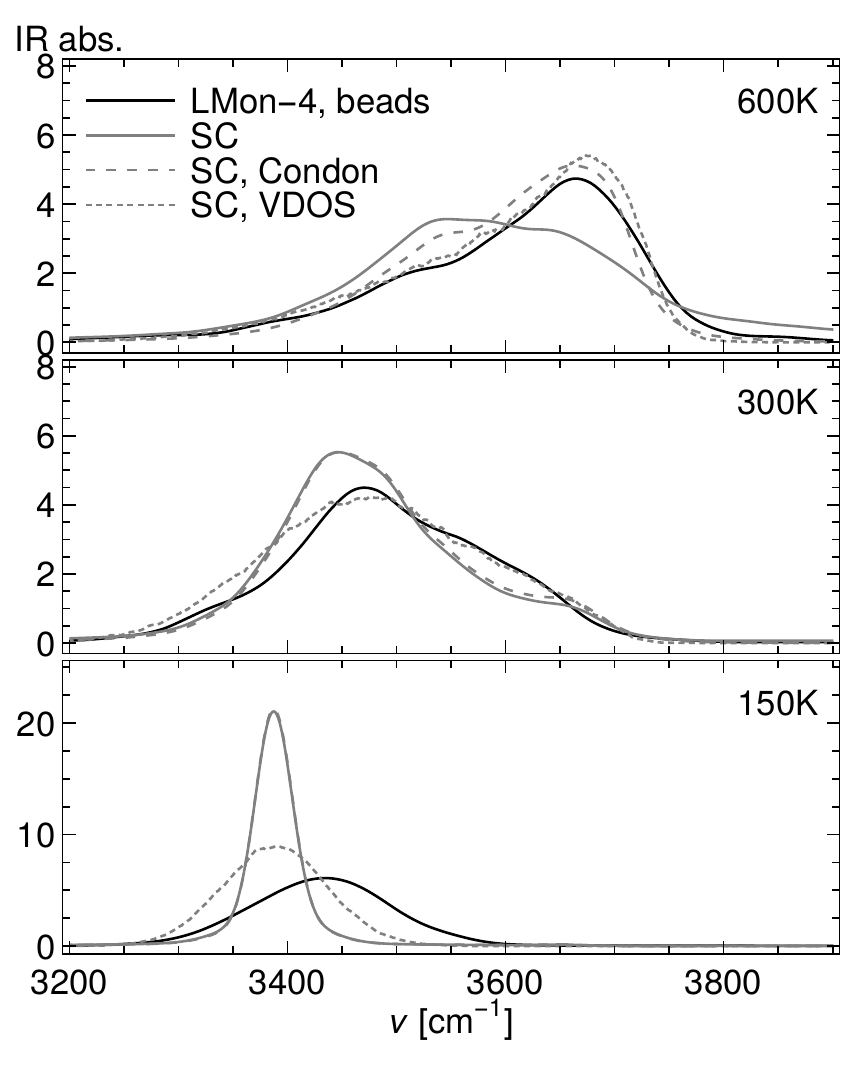}
\end{center}
\caption{Comparison between LMon-4 spectra and SC line shapes for the OH stretch in HOD:D$_2$O at different temperatures. The SC lineshape including non-Condon corrections is compared with the vibrational density of states (VDOS) and with the SC lineshape \emph{without} non-Condon corrections (Condon). Note that at 150K the Condon and non-Condon profiles are indistinguishable.}
\end{figure}

\section{Details of SC calculations}
Following Ref.~\cite{mukamel1999principles}, all SC line shapes were calculated using molecular configurations extracted from 10 CMD trajectories of 50 ps. Briefly, for any instantaneous CMD configuration of the HOD:D$_2$O system, the OH vibrational frequencies and wave functions of both ground and first-excited states were calculated by solving the Schr\"{o}odinger equation for the corresponding one dimensional oscillator using the Numerov method\cite{Noumerov_1924}. The underlying OH potential energy curve was obtained by stretching the OH bond of HOD while keeping the positions of all other atoms fixed. The associated vibrational transition moment ($\mu_{10}$) was calculated within the same computational scheme from the numerical integration of the product of the molecular dipole moment and the appropriate vibrational wave functions. 
The non-Condon OH line shape for the 1-0 vibrational transition w classical limit was calculated as
\begin{equation}
I^\text{NC}(\omega) = {{1}\over{2\pi}} \int_{-\infty}^{\infty} dt e^{-i\omega t}
\langle {\bf \mu}_{10}(0)\cdot {\bf \mu}_{10}(t) e^{i\int_0^t d\tau \omega_{10}(\tau)}\rangle
\end{equation}
where $\mu_{10}$(t) is the vibrational transition moment associated to the transition between the ground and excited state of the OH bond, $\omega_{10}$(t) is the instantaneous vibrational frequency at time t, and the brackets denote a classical equilibrium average. Within the Condon approximation, it is then assumed that the magnitude of the vibrational transition moment is independent of the surrounding environment, which leads to 
\begin{equation}
I^\text{C}(\omega) \propto  \int_{-\infty}^{\infty} dt e^{-i\omega t}
\langle e^{i\int_0^t d\tau \omega_{10}(\tau)}\rangle.
\end{equation}

\begin{figure}[tbhp]
\begin{center}
\includegraphics[width=0.7\columnwidth]{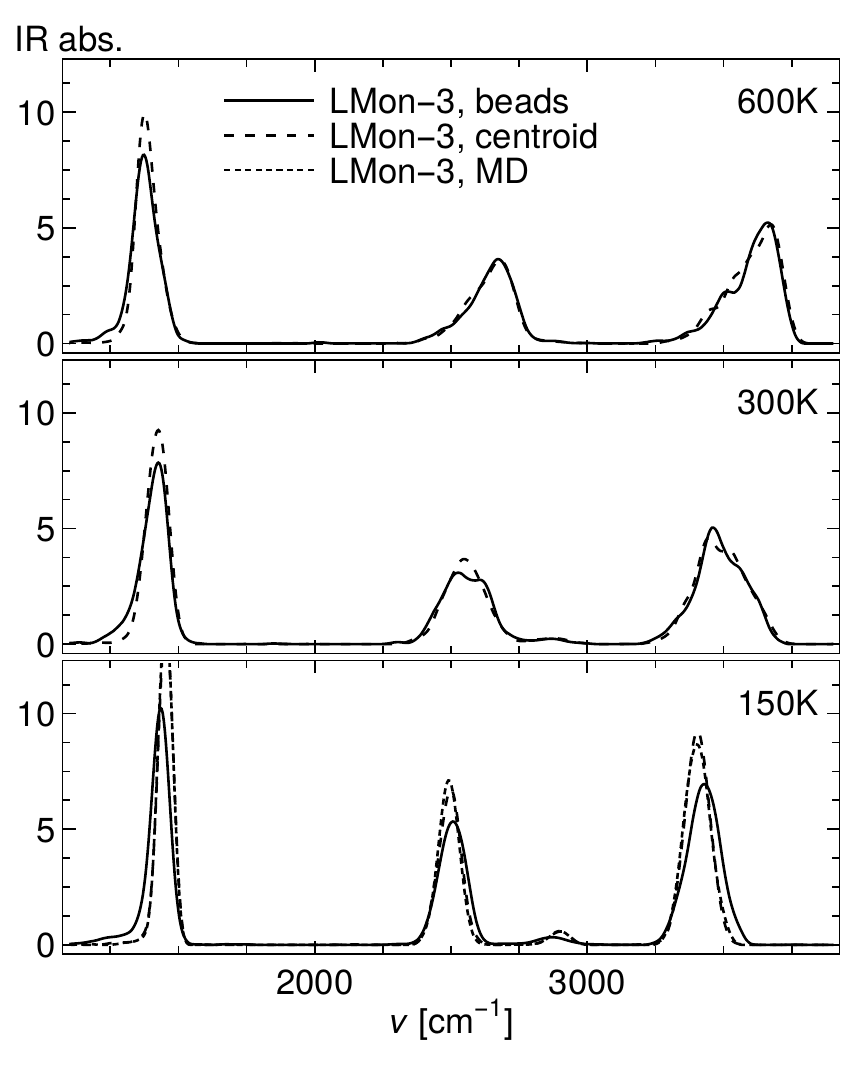}
\end{center}
\caption{Dependence of LMon-3 on the atomic configuration of HOD in D$_2$O in the three thermodynamic conditions considered in the manuscript. We compare configurations coming from the beads and from the centroid of PIMD simulations, and from a classical MD simulation. The largest differences are seen for ice at 150K.}
\end{figure}

The VDOS was obtained simply from making a histogram of the frequencies calculated from all snapshots, but without weighting by the dipole transition moment.

\section{LMon dependence on the finite-temperature structure}

In the following we study the dependence of the LMon-3 spectra to the use of beads or centroid configurations from PIMD simulations, or configurations coming form a classical MD simulation. 
The differences are mainly seen at 150K. For OH band, the centroid result is 25 cm$^{-1}$ red-shifted with respect to the result for the bead configurations. For the OD band, centroid-derived spectrum has a 10 cm$^{-1}$ red-shift if the center of the band is considered. For the bend, centroid has a 18 cm$^{-1}$ blue-shift. 
For ice HOD:D$_2$O at 150K also the result using classical MD configurations is shown, being quite similar to the one coming from the centroid configurations, which highlights the close-to-classical nature of the centroid in PIMD simulations.

\begin{figure}[tbhp]
\begin{center}
\includegraphics[width=0.7\columnwidth]{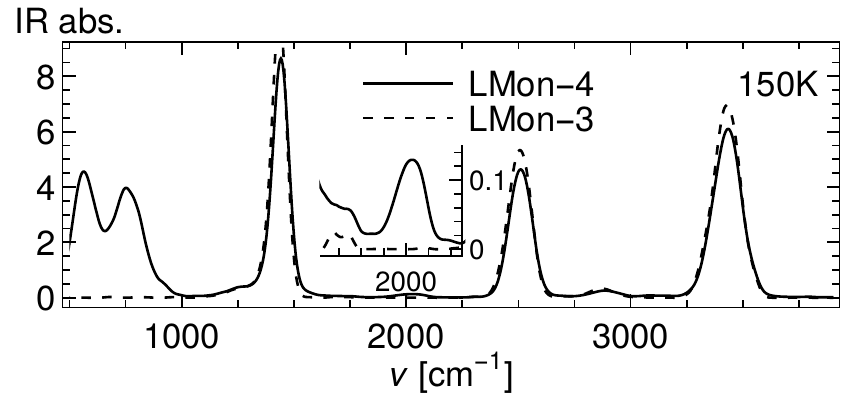}
\end{center}
\caption{Differences in the spectra obtained from the LMon-3 (dashed) and LMon-4 (full) methods of HOD in D$_2$O for Ice at 150K. The inset shows a magnification of the combination band of bend and libration that only LMon-4 captures.}
\end{figure}

\begin{figure}[tbhp]
\begin{center}
\includegraphics[width=0.7\columnwidth]{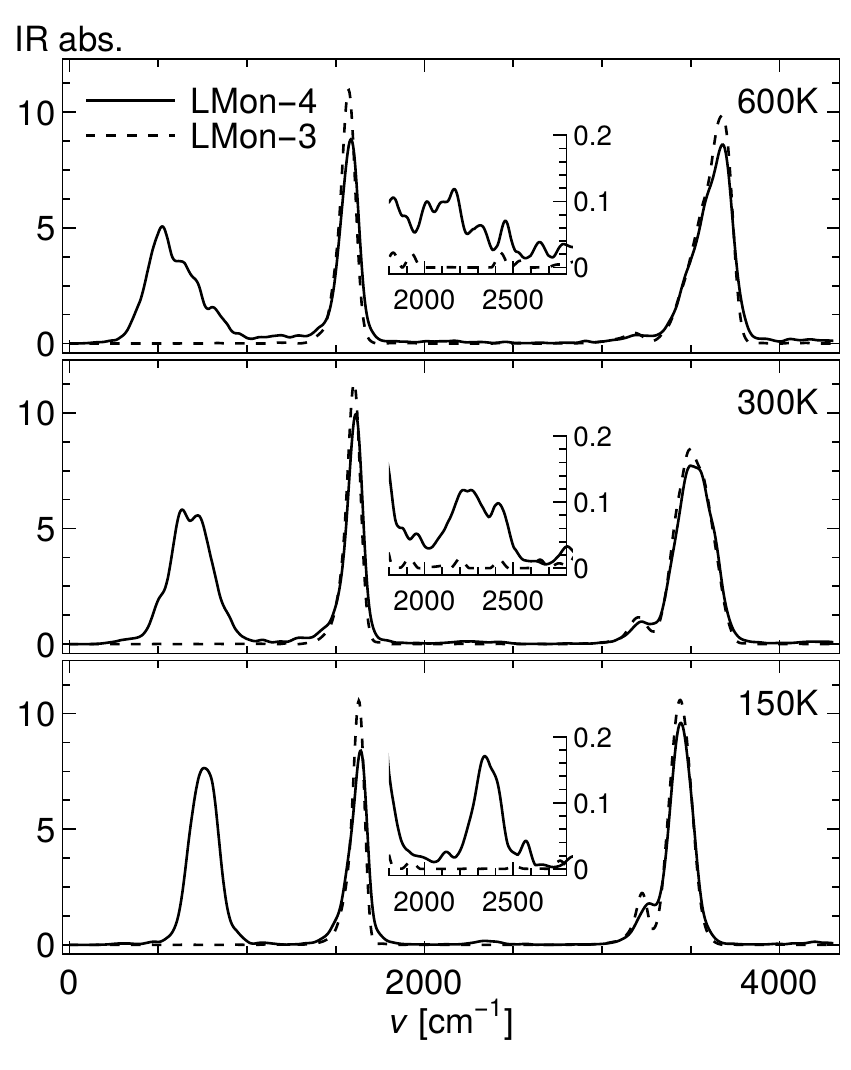}
\end{center}
\caption{Differences in the spectra obtained from the LMon-3 (dashed) and LMon-4 (full) methods of pure light water. The insets show a magnification of the combination band of bend and libration that only LMon-4 captures. }
\end{figure}

\section{LMon dependence on the quantum subspace size}

We here study the differences between a LMon-3 and a LMon-4 (in the "effective LMon-6" approximation) calculations for the IR spectrum of neat water (H$_2$O) and HOD in D$_2$O.
First, the intramolecular bands from LMon-4 and LMon-3 are nearly the same. This implies LMon-3 is an effective and accurate method to compute intramolecular bands of the vibrational spectra of condensed-phase matter.
Second, LMon-4 is able to capture also intermolecular bands. For H$_2$O, one of them is a librational band at 400-1000 cm$^{-1}$ and the other is a combination band of bend and libration at 2100-2300 cm$^{-1}$. 
For HOD, the librational band is in the range of 400-900 cm$^{-1}$ and the combination band appears at around 2000 cm$^{-1}$. Expanding the subspace to include a intermolecular mode is therefore an important step. However, LMon-4 is still far from the optimal real LMon-6 (or the ultimate LMon-9), which are much more computationally expensive. We thus believe that the intermolecular bands obtained from separate LMon-4 calculations are considered to be less accurate than the intramolecular bands. 

\end{document}